\RequirePackage{fix-cm}
\documentclass[smallextended]{svjour3}       % onecolumn (second format)
\smartqed  % flush right qed marks, e.g. at end of proof
\usepackage{graphicx}
\usepackage [latin1]{inputenc}
\usepackage{natbib} 
\newcommand{\micron}{\mu m}
\begin{document}

\title{New active galactic nuclei science cases with interferometry %\thanks{Grants or other notes
}
\subtitle{An incomplete preview}

\titlerunning{AGN with interferometry}        % if too long for running head

\author{Sebastian F. H\"onig         \and
 Almudena Alonso Herrero        \and
 Poshak Gandhi        \and
 Makoto Kishimoto        \and
 J\"org-Uwe Pott        \and
 Cristina Ramos Almeida        \and
 Jean Surdej        \and
 Konrad R. W. Tristram
}


\institute{S.F. H\"onig \at
              Department of Physics \& Astronomy, University of Southampton, SO17 1BJ\\
              \email{S.Hoenig@soton.ac.uk}           %  \\
           \and
A. Alonso Herrero \at
              Centro de Astrobiología (CAB, CSIC-INTA), ESAC Campus, E-28692 Villanueva de la Cañada, Madrid, Spain
           \and
P. Gandhi \at
              Department of Physics \& Astronomy, University of Southampton, SO17 1BJ
           \and
M. Kishimoto \at
              Kyoto Sangyo University, Kyoto 603-8555, Japan
           \and
J.-U. Pott \at
              Max-Planck Institut f\"ur Astronomie, K\"onigstuhl 17, D-69117 Heidelberg, Germany
           \and
C. Ramos Almeida \at
              Instituto de Astrofisica de Canarias (IAC), E-38205 La Laguna, Tenerife, Spain
	  \and
J. Surdej \at
              Space sciences, Technologies and Astrophysics Research (STAR) Institute, Universit\'e
 de Li\`ege, 19c All\'ee du Six Ao\^ut, B-4000 Li\`ege, Belgium
	  \and
K.R.W. Tristram \at
              European Southern Observatory, Casilla 19001, Santiago 19, Chile}

\date{Received: date / Accepted: date}

\maketitle

\begin{abstract}
Infrared (IR) interferometry has made widely recognised contributions to the way we look at the dusty environment of supermassive black holes on parsec scales. It finally provided direct evidence for orientation-dependent unification of active galaxies, however it also showed that the classical ``torus'' picture is oversimplified. New scientific opportunities for AGN have been suggested, and will soon be carried out, focusing on the dynamical aspects of spectrally and spatially resolved interferometry, as well as the potential to employ interferometry for cosmology. This will open interferometry to new scientific communities.
\end{abstract}

\section{The past ten years: The dusty environment of active galactic nuclei}\label{sec:1}

Long-baseline infrared (IR) interferometry is a rather young technique for extragalactic science. The key challenge for this field of science is the combination of sensitivity and baseline length of current interferometers. Most compact extragalactic targets are faint for interferometry purposes and require the light collecting power of at least 8\,m class telescopes.  

The first successful observations of the dust around supermassive black holes in active galactic nuclei have been reported in the mid-2000s when interferometers with 8-10\,m class telescopes became available \citep[e.g.][]{Swa03,Wit04,Jaf04}. These studies targeted the two IR-brightest AGN in the sky: NGC4151 in the northern hemisphere with the Keck interferometer and NGC1068 in the south with the VLTI. Yet even with those large-scale facilities, observing new objects remains challenging up to today since the field is mostly sensitivity limited. Nevertheless, with progress on the instrumentation side, both on hardware and software, a sample of up to 50 nearby AGN has been successfully observed by now \citep[for a recent status update see][]{Bur16}.

The major focus of the first decade of extragalactic long-baseline interferometry was the study of the dusty environment of actively accreting supermassive black holes in the centres of galaxies. One of the first scientific advancement that spun out of interferometry was the realisation that the dust around the supermassive black hole is inhomogeneously distributed \citep[i.e.``clumpy''; see e.g.][]{Jaf04,Sch05,Hon06,Tri07}. This confirmed a prediction that had been made mostly based on theoretical considerations and indirect evidence \citep[e.g.]{Kro88,Nen02}.

The two-telescope mid-IR beam combiner MIDI at the VLTI was responsible for the majority of all interferometric AGN observations. It provided spectrally-resolved visibility and chromatic phase information in the $8-13\,\micron$ range, tracing dust with typical temperatures of about 200-300\,K. The wavelength range is close to the peak in the infrared emission from AGN (at about $20-30\,\micron$), thus tracing the dust distribution where most of the reprocessing occurs. With the help of the growing sample of AGN observed with MIDI, it was realised that the dusty environment among different objects is quite diverse \citep{Kis11b,Bur13}. This diversity may be influenced by accretion rate and/or luminosity of the AGN, but a clear answer to this is still pending. 

\begin{figure*}
\begin{center}
  \includegraphics[width=0.7\textwidth]{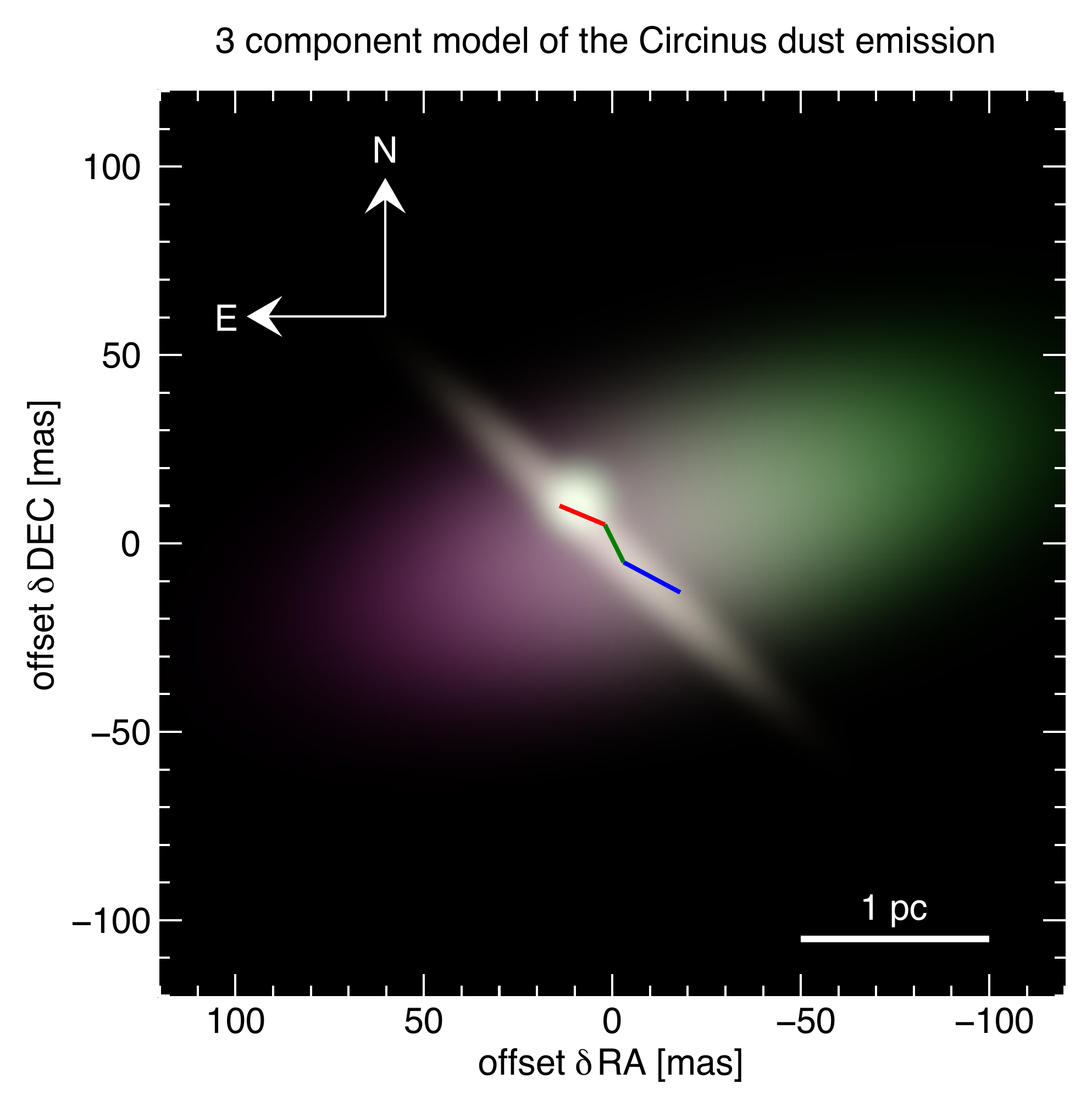}
\end{center}
\caption{Reconstructed model image of the mid-infrared emission from the centre of the Circinus galaxy. The
VLTI/MIDI reveal that about 80\% of the observed emission emerges from the polar region of the AGN, extending
from the parsec scales to tens of parsecs. A weaker component, accounting for most of the remaining flux, is
interpreted as a geometrically thin disk in the equatorial plane of the AGN. This configuration is not compatible with
current torus models. From \citet{Tri14} in Astronomy \& Astrophysics.}
\label{fig:1}       % Give a unique label
\end{figure*}

More recently, detailed mid-IR interferometry of the brightest targets indicated that the classical torus picture is probably too simplistic. Instead, the infrared emission was observed to consist of a strong emission component along the direction of gas outflow approximately perpendicular to what was considered the torus, and a geometrically thin equatorial emission component \citep[][see Fig.~\ref{fig:1} for the Circinus galaxy as an example] {Hon12,Hon13,Tri14,Lop16}. The observed features are not naturally recovered in classical torus models. As a remedy, a new disk+wind model was introduced that is able to simultaneously reproduce IR SED and interferometry of AGN \citep{Hon17}. If confirmed, this represents a significant shift of paradigm in the field of AGN. Observational and theoretical efforts into this direction are ongoing \citep[e.g.][]{Wad15,Asm16}.

\section{New scientific pathways for the next decade} \label{sec:2}

Given the strong limits set on extragalactic interferometry by sensitivity, it might be expected that any new scientific fields will be opened on interferometers with 8\,m class telescopes. Given the unavailability of Keck for interferometry at the moment, this leaves us with the VLTI. However, a significant factor in sensitivity is rigidity of the overall observatory and high transmission, starting at the telescopes and going down to the optical elements that carry the light through the delay line infrastructure into the instruments. Thus, simplicity can win over size. As an example, ongoing efforts by M. Kishimoto and collaborators as well as the observatory staff led to the first detection of AGN fringes with the 1\,m CHARA array.

For the sake of this preview (as opposed to a review), new scientific projects using the sensitivity and resolution limit of the VLTI are presented. The focus is specifically on MATISSE \citep{Lop14} and GRAVITY \citep{Eis11}, which are new four-telescope interferometric instruments in the mid- and near-infrared, respectively. The step to four telescopes will provide phase information, which is crucial for image reproduction, and enables "super resolution" (see below). In addition, MATISSE will open the atmospheric $L$ and $M$ bands to interferometry, which provides a wealth of new opportunities for science with emission lines or intermediate-hot dust. The following outlines some near- and mid-term science cases that are being actively studied and pushed forward.

\subsection{Near-term science: Toward a new paradigm for the dust distribution around AGN}

A natural follow-on from the past success of interferometry is pinning down the dust distribution around AGN. The strong and often dominating mid-IR polar emission features in AGN were only discovered towards the end of the lifetime of MIDI. Key open questions that will be addressed from the start of operations of MATISSE will be: Why do we see a two-phase structure? What is the physical origin? How do the parsec scale dusty wind features fit into the large-scale AGN winds and black hole feedback? Answering these questions is fundamental to our understanding of the physical processes that regulate accretion onto black holes and produce the diversity in AGN classes. These efforts will require targeted, detailed interferometry and further pushes in theory towards full radiative (magneto-)hydrodynamical simulations of the dusty environment.

While the 8\,m telescopes on the VLTI will not provide additional resolution beyond what has been achieved by MIDI, the new piece of information will be phases, combined with prompt coverage of a good part of the uv plane. Each interferometric observation will deliver six visibility points and three closure phases. In one night, one may expect to fill the uv plane with 24 independent visibility points and 18 closure phases. The same level of visibility data (without closure phases) is available only for very few AGN at the moment (NGC1068, Circinus, NGC3783, NGC424) and required easily a dozen nights per object distributed over many years to collect. Just by the amount of additional data, MATISSE has the potential to become transformational for AGN science. 

For this to happen, MATISSE has to achieve at least the same sensitivity as MIDI. The latter was an optically simple instrument as compared to MATISSE, thus minimising possible transmission losses. To overcome the risks of lower sensitivity, it has been suggested to use GRAVITY as a near-infrared fringe tracker for MATISSE (``GRAV4MAT''). This would stabilise fringes in the mid-IR and allow for longer exposure times. It seems that this is a crucial piece to exploit the full potential of MATISSE for extragalactic science.

\subsection{Near-term science: Unveiling the black hole in the centre of the Milkyway}

GRAVITY is now offered for observations on the VLTI. It has been designed for specific science cases related to the supermassive black hole in the centre of our own Galaxy\footnote{The notion that the black hole in the centre of the Galaxy constitutes an extragalactic science case may seem a non-squitur. However, the centre of the Galaxy serves as a proxy -- the nearest example -- for galactic centres in general.}. The principle science cases use the astrometric mode of GRAVITY where the position/motion of a target object is tracked relative to a steady reference source. This way it will be possible to trace the motion of matter orbiting the black hole close to the event horizon and use the motion of stars close to the black hole to measure quadrupole moment of the black hole, ultimately probing general relativity in the strong gravity regime.

The dynamic processes even around a "nearby" supermassive black hole as the one in our Galaxy occur on $\mu$-arcsecond scales, which is beyond the classical resolution limit of a 100-200\,m interferometry when working with visibilities. Thus, GRAVITY's astrometric mode will use the phase information. Small phase shifts between reference and target can be converted into a projected separation. Over time, the change in phase will allow to map the motion of the target object. 

\subsection{Mid-term science: AGN as a standard ruler}

Constraining the parameters of the standard $\Lambda$CDM cosmological model has become one of the top adventures globally in astronomy. Of particular interest are the equation-of-state of dark energy, expressed as $w_0$ and $w_a$, and the local expansion rate, i.e. the Hubble constant $H_0$. In recent years, a tension has emerged between $H_0$ measured from the local distance ladder and the Hubble constant inferred from the cosmological microwave background \citep[e.g.][]{Rie16}. While this may point towards new physics, trivial reasons of such a discrepancy can be unknown systematic effects in either the local or CMB determination of $H_0$. Thus, an independent test is warranted to probe if such systematic errors are plausible.

\begin{figure*}
  \includegraphics[width=\textwidth]{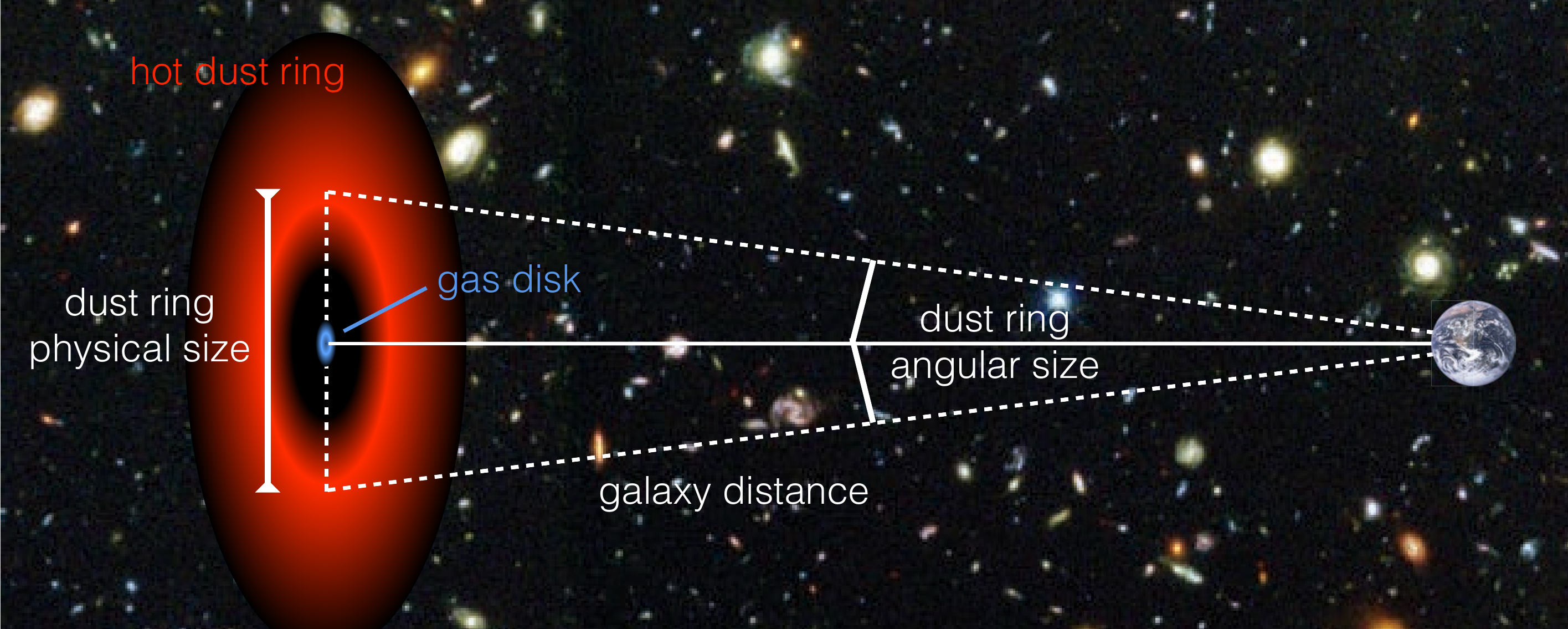}
\caption{Illustration of the AGN dust parallax method to measure direct distances to AGN. Near-IR interferometry
measures the angular size of the hot dust ring around the AGN while reverberation mapping can determine the
physical size of the structure. The distance can be inferred from trigonometry.}
\label{fig:2}       % Give a unique label
\end{figure*}

It was recently suggested that the combination of AGN near-IR time-delay measurements and interferometry can be used to measure geometric distances to extragalactic targets \citep[][see Fig.~\ref{fig:2} for an illustration of the method]{Hon14}. From those distances, $H_0$ can be directly inferred from trigonometry without the use of the distance ladder, providing the sought-for test. A sizeable fraction of the brightest AGN is located in the Hubble flow at distances $>100 Mpc$, which is necessary to have the galaxy redshift being dominated by cosmic expansion rather than peculiar motions. 

This science case opens interferometry to a new scientific community. The major challenge concerns the accuracy of the near-IR size measurements. Since AGN hot dust sizes in the near-IR are of the order 1\,milliarcsecond or less, visibilities for any $K$-band instrument using the VLTI 8\,m telescopes will be of the order 0.9. Therefore, high S/N is required, which means stable fringe tracking and long exposures. Of similar importance is calibration accuracy. On AO-assisted telescopes, the quality of the AO correction influences the shape/Stehl factor of the beam injected into the interferometric instrument. The near-IR emission of unobscured AGN is a combination of the hot dust emission and star light that is extended on scales resolved by the individual telescopes. This results in worse AO performance for AGN than for reference stars, potentially leading to lower instrumental visibilities and in turn to a size estimate that is too large. Thus, understanding of calibration of GRAVITY will be very important to enable the direct distance measurements.

\subsection{Mid-term science goal: Dynamical black hole mass measurements in the local universe and beyond}

The most accurate and direct way of measuring masses of supermassive black holes are kinematically mapping stars or gas in its sphere of influence. For masses of $10^6-10^8\,M_\odot$, the resolution of a single 8-10\,m telescope is sufficient out to distance of about 30 Mpc. Beyond this, interferometric resolution is required. Measuring these masses are an important prerequisite to understand co-evolution of galaxies and their black holes. 

A way to overcome the resolution limit of individual telescopes is resolving the sphere of influence in the time domain. The variability of the accretion disk emission is mirrored in the emission of surrounding gas clouds that act as a reprocessor. Therefore, when monitoring the optical continuum emission from the AGN and line emission from these clouds, both light curves will look very similar in shape, however with the emission line variability lagging the continuum emission by several days to weeks, corresponding to the light travel time between the accretion disk and the gas clouds. In the ideal case, lags can be estimated for different parts in the emission line, hence connecting projected distance from the AGN with velocity. The reverberation mapping technique takes the input of emission line lags and line widths and converts them into a black hole mass via the virial theorem. However, complicated kinematics of inflows and outflows and projection effects cause significant systematic uncertainties. 

The spectral window of GRAVITY ($2.05-2.45\,\micron$) contains either the Br$\gamma$ or Pa$\alpha$ broad emission lines, depending on redshift of the object. Given GRAVITY's spectral resolution of R$\sim$4000 provides sufficient spectral resolution to measure the size of the broad-line region depending on gas velocity. In combination with the shape of the spectral line and the visibility and phases accross the line, it will be possible to disentangle geometry and kinematics, leading to accurate black hole mass measurements. Sensitivity is the biggest challenge for this science case given that the spectral dispersion dilutes the signal. Therefore, stable fringe tracking will be required to achieve statistical uncertainties of black hole masses comparable to reverberation mapping.

\section{Summary and conclusions}

New capabilities in infrared interferometry have the power to widen the interest in very high angular resolution observations beyond currently covered extragalactic fields of research. Specifically, tests for general relativity in the strong gravity regime, cosmological applications, and measurement of black hole masses will appeal to new scientific communities. A key to success is sensitivity. Stable fringe tracking, reliable beam injection, and low transmission losses should be considered a priority when upgrading the VLTI infrastructure to facilitate a wide range of extragalactic science.

%\begin{acknowledgements}
SFH acknowledges support from the Horizon 2020 ERC Starting Grant DUST-IN-THE-WIND (ERC-2015-StG-677117).
%\end{acknowledgements}



\end{document}